IAC-24-A7.2.89135

## The AGILE space mission: an Italian success story and its legacy for future space astronomy

**INAF: C. Pittori[a,b]\*, M. Tavani[c,d], A. Argan[e], A. Bulgarelli[f]**

**INFN: G. Barbiellini[g]**

**ASI: L. Salotti[h], E. Tommasi[h], E. Cavazzuti[h], F. D'Amico[h], G. Valentini[h], M. Jahjah[h], B. Negri[h]**

**OHB: G. Annoni[i]**

**on behalf of the AGILE Collaboration**

[a] *INAF/OAR Roma, Via di Frascati 33, I-00078 Monte Porzio Catone, Italy, carlotta.pittori@inaf.it*
[b] *ASI Space Science Data Center (ASI-SSDC), Via del Politecnico snc, 00133 Roma, Italy*
[c] *INAF/IAPS Roma, via del Fosso del Cavaliere 100, I-00133 Roma, Italy*
[d] *Università degli Studi di Roma 'Tor Vergata', via della Ricerca Scientifica 1, I-00133 Roma, Italy*
[e] *INAF-Headquarter, Viale del Parco Mellini 84, 00136 Rome, Italy*
[f] *INAF/OAS Bologna, Via P. Gobetti 93/3, 40129 Bologna, Italy*
[g] *INFN Trieste, Padriciano 99, I-34012 Trieste, Italy*
[h] *Agenzia Spaziale Italiana (ASI), via del Politecnico snc, I-00133 Roma, Italy*
[i] *OHB Italia SpA, Via Gallarate, 150, 20151 Milano, Italy*

\* Corresponding Author

### Abstract

AGILE (Astrorivelatore Gamma ad Immagini LEggero) has been a unique and hugely successful mission of Italian Space Agency (ASI), built and operated with the programmatic and technical support of the National Institute for Astrophysics (INAF), the National Institute for Nuclear Physics (INFN), several universities and industrial industries. During almost 17 years of observations in orbit (from April 23, 2007, to January 18, 2024), AGILE contributed to high-energy astrophysics and terrestrial physics with many discoveries and detections. Two co-aligned X- and gamma-ray detectors, a silicon-strip-based tracker, a wide field of view gamma-ray imager and the fast-reaction ground segment were the AGILE innovative solutions with respect to the previous generation of gamma-ray satellites. With the AGILE's re-entry, the in-orbit operational phase ends, but a new phase of scientific work on the satellite legacy data archive opens: AGILE may still hold future surprises.

### Acronyms/Abbreviations

Agenzia Spaziale Italiana (ASI)
Findable Accessible Interoperable Reusable (FAIR)
Fast Radio Burst (FRB)
Gamma-ray Imaging Detector (GRID)
Gamma-ray Burst (GRB)
Gravitational Waves (GW)
Indian Space Research Organisation (ISRO)
Istituto Nazionale di Astrofisica (INAF)
Istituto Nazionale di Fisica Nucleare (INFN)
Space Science Data Center (SSDC)
Terrestrial Gamma-ray Flashes (TGF)

## 1. Introduction

AGILE (Astrorivelatore Gamma ad Immagini LEggero) is an Italian scientific space mission for high energy astrophysics funded by the Italian Space Agency (ASI) with scientific and programmatic participation by INAF, INFN, several Italian universities and industrial contractors [1]. AGILE is a Principal Investigator (PI)-led mission of ASI with scientific leadership by INAF and Co-PI-ship by INFN. Operational activities were carried out in agreement with the ASI Mission Director.

The AGILE satellite, designed for a nominal operative life of only two years, was launched on April 23, 2007, and ceased observations on January 18, 2024, after almost 17 years of successful in-orbit operations. The satellite re-entered the Earth's atmosphere on 14 February 2024, due to the natural decay of its low Earth orbit. AGILE was successfully launched from the Indian base of Sriharikota (Fig. 1), and it was inserted in an equatorial low Earth orbit with an inclination of about 2.5 degrees and initial average altitude of about 500 km.

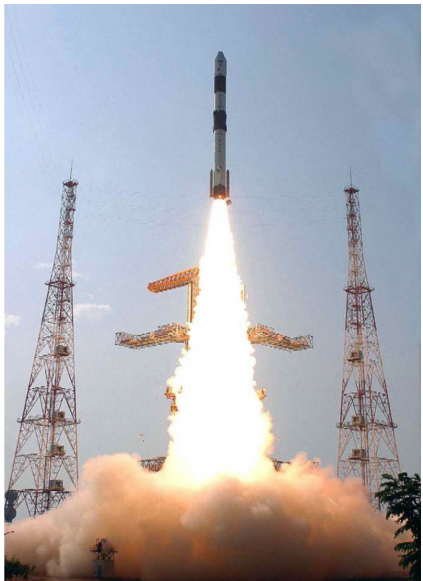

Fig.1. The successful AGILE launch on April 23, 2007 by the Indian PSLV-C8 rocket.

The spacecraft worked autonomously without ground station control for most of its orbital period of about 100 minutes, therefore the data generated by the detectors were stored on board in a mass memory to be downloaded to the ASI Malindi ground station during the first available visibility period. During ground station contact, AGILE raw telemetry level-0 data were periodically downlinked to the ASI Malindi ground station in Kenya and transmitted, through the fast ASINET network provided by ASI, first to the Telespazio Mission Control Center at Fucino in Italy and then to the AGILE Data Center, which is part of the ASI multi-mission Space Science Data Center (SSDC, previously known as ASDC), about 5 minutes after the end of each contact downlink (see Fig. 2).

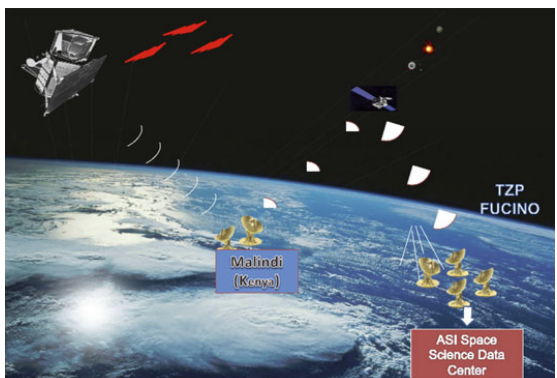

Fig. 2. The AGILE data flow (a pictorial view).

## 2. The AGILE satellite and payload

The AGILE satellite and payload are schematically illustrated in Fig. 3.

The main satellite technical data were the following:
- Mass at launch: 350 kg
- Satellite dimensions: (1,7 x 2 x 0,8) m$^3$
- Electrical Power: 200 W (average)
- Sun pointing attitude, one fixed solar panel
- Attitude knowledge: 1 arcmin
- On board autonomy of 3 days without ground contacts

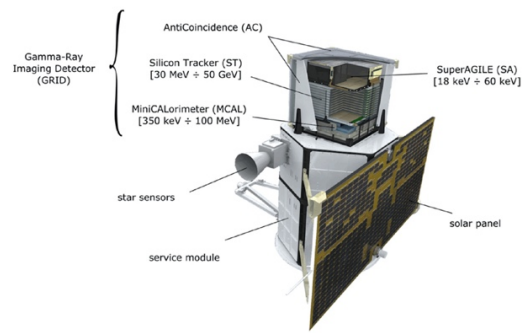

Fig. 3. Schematic view of the AGILE satellite and the payload scientific instruments.

The AGILE very innovative payload combined for the first time a gamma-ray imager based on solid-state silicon technology and a hard X-ray imager. The whole payload was a cube of about 60 cm size and of weight near 100 kg. The AGILE scientific payload consists of four instruments with independent detection capability (see Fig. 3):

1. the gamma-ray imager detector (GRID), consists of a silicon–tungsten tracker, a cesium iodide mini calorimeter, and an anticoincidence system, working together to detect gamma-rays in the energy range 30 MeV- 30 GeV with a large field of view ($\sim$ 2.5 sr);
2. a mini calorimeter (MCAL), sensitive in the energy range 300 keV to 100MeV that works both as a subsystem for the GRID and as an autonomous detector for transient events;
3. a hard X-ray imager on top (Super-AGILE), sensitive in the energy range 18–60 keV with a field of view of $\sim$ 1 sr;
4. an anticoincidence system (ACS) of segmented plastic scintillators surrounding the AGILE detectors, that works both as a subsystem for the GRID and as an autonomous detector.

A data handling (DH) system with sophisticated processing capabilities completes the instrument together with a power-supply unit [2].

More details and related references about the AGILE instruments are given in the following subsections.

### 2.1 The AGILE silicon tracker

The core instrument of the AGILE GRID is a silicon-tungsten tracker developed in the Trieste INFN laboratories [3,4], with the contribution of LABEN (now Thales Alenia Space - Italia) and Oerlikon Contraves (now Rheinmetall Italia), and assembled at the Mipot facility in Cormons (Gorizia), Italy (see Fig. 4).

The tracker gamma-ray detection mode is based on the photon conversion process into electron-positron pairs. The AGILE tracker is made of 12 layers with a distance of 1.9 cm between them, and the first ten layers are equipped with a thin tungsten converter layer of 245μm (0.07 X0) each. The trajectories of the charged particles are tracked by silicon strip detectors arranged to provide two orthogonal measurements for each track element.

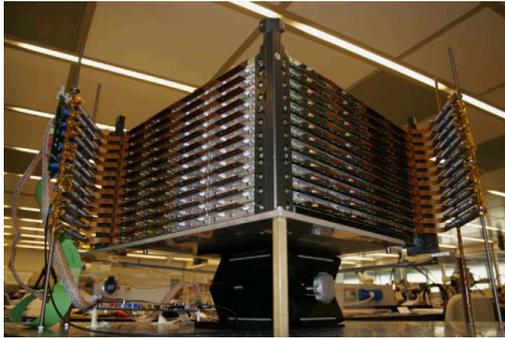

Fig. 4. The AGILE silicon tracker developed in the Trieste INFN laboratories in the Mipot facility in Cormons (Italy) during testing (June 2005).

A key characteristic of the AGILE tracker readout system is the capability of detecting the energy released by charged particles in the silicon microstrips. The microstrips have a pitch of 121μm and half of them are read by an alternating readout system. This combination of digital and analog information is unique to the AGILE tracker which results in an excellent positional resolution, better than 40μm for a wide range of particle incidence angles [5].

### 2.2 The AGILE MCAL

The AGILE mini-calorimeter (MCAL) is a non-imaging detector with a $4\pi$ sr acceptance and consists of 30 cesium iodide (CsI) bars placed in two orthogonal layers for a total (on-axis) radiation length of 1.5 X0. The MCAL instrument was designed and developed at the Bologna section of INAF [6,7]. The system's responsibility for its realisation was entrusted to LABEN, now Thales Alenia Space - Italia (see Fig. 5). The calibration at the integrated satellite level was carried out at the Carlo Gavazzi Space clean room in Tortona, now OHB Italia.

MCAL is sensitive in the range 300 keV–100 MeV and gives a measure the energy of photons converted in the tracker, but it can also work independently to detect GRBs, TGFs and other impulsive gamma-ray events thanks to a special standalone "burst search algorithm", over a broad range of trigger timescales ranging from sub-milliseconds to many seconds.

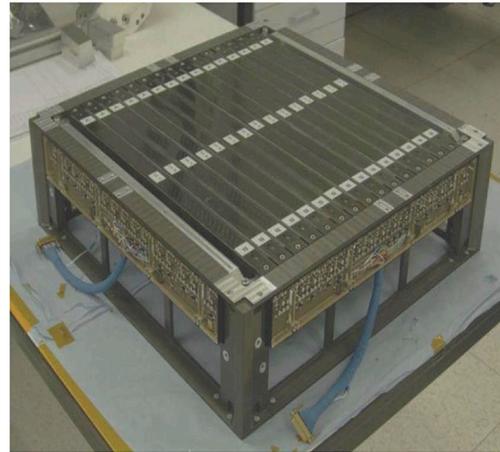

Fig. 5. The AGILE MCAL conceived and developed at the INAF Bologna section, shown during an early stage of integration in the Thales Alenia Space laboratories in Milan (February 2005).

### 2.3 Super-AGILE

Super-AGILE is a hard X-ray imager put on top of the gamma-ray tracker, made of four square Silicon detectors (19x19 cm2 each) plus an ultra-light coded mask system supporting a Tungsten mask placed at a distance of 14 cm from the Silicon detectors. Super-AGILE is sensitive in the 18–60 keV band and was developed at the INAF Rome section [8] (see Fig. 6). The integration of SuperAGILE was largely conducted at Mipot facility in Cormons (Italy), with the direct participation of the scientific team.

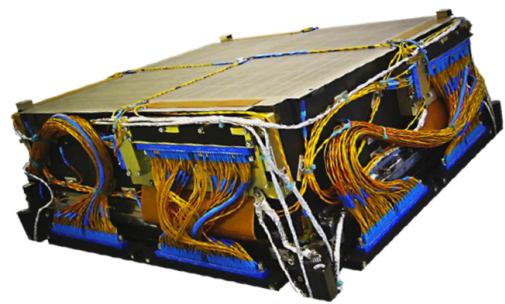

Fig. 6. The Super-AGILE detector developed at the INAF-IAPS Rome section, shown during laboratory test measurements (March 2005).

*2.4 The AGILE ACS*

The ACS was designed in INAF Milano and it comprises five independent plastic scintillation panels (four lateral and one on top) [9] (see Fig. 7). The anticoincidence (AC) primary role is to reject charged background particles, but they are also sensitive to hard X-ray photons in the 50 – 200 keV energy range.

Each AC lateral side is segmented into three independent panels of plastic scintillators (0.6 cm thick), this segmentation is essential for providing information on the direction reconstruction of photons at large off-axis angles and thus for achieving the very large field of view of the AGILE-GRID imager.

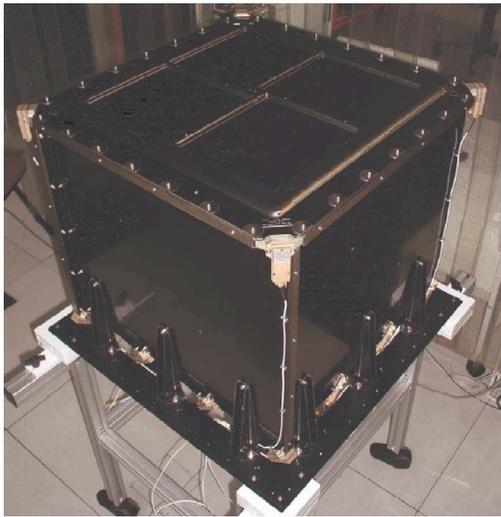

Fig. 7. The AGILE anticoincidence detector flight unit.

## 3. The ASI management for the AGILE mission

The AGILE space mission started in 1997 as a project selected in the frame of the "Small Mission" program of ASI. ASI led and funded the whole mission, including the payload, as well as the ground segment and launch campaign. Since the S/C launch in April 2007, the ASI ground segment management team was led by an ASI Mission Director that coordinated the program activities and was responsible of all the technical and management decisions. Since its start, different ASI staff members were assigned the AGILE Mission Director role.

In December 2007, at the completion of the commissioning phase, AGILE entered the nominal scientific phase organized in annual cycles. In this perspective, ASI periodically issued an "Announcements of Opportunity" for the Italian and international scientific community, receiving "Observation Proposals" to be examined by the AGILE Data Allocation Committee. The official scientific observation plan was approved by the AGILE Mission Board (AMB), who was responsible for assessing, close to the expiry of each cycle, the actual state of health of AGILE through the analysis of the functioning of its main subsystems, in such a way as to guarantee the operability of the satellite for the period concerning a possible new extension, and consequently a new cycle. Following the AGILE Mission Boards positive assessment, the mission was extended year by year. The last mission extension started in July 2023 and ended with the re-entry of the satellite in the atmosphere in February 2024. In fact, on 10/02/2023, in view of the expiration of the cycle 2022-2023, the AGILE Mission Board met and expressed a favourable opinion to extend the mission by another 12 months in the period 2023-2024. This assessment was based on the confirmation of the scientific validity and excellence of the mission for both astrophysics and terrestrial physics. The AMB also confirmed that the scientific payload was still in nominal operation and that gamma ray observations were still nominal. Concerning terrestrial physics, the AMB also confirmed the significant increase in sensitivity of the AGILE Mini-Calorimeter (MCAL) payload with very interesting results.

However, the monitoring of the AGILE orbital parameters started to display a significant decay of the satellite altitude in summer 2023. The orbital decay was discussed in several meetings between ASI, the industrial contractors (OHB and Telespazio) and the Centre for Space Situation Awareness of the Italian Air Force of Poggio Renatico/FE in order to decide on the final procedures and date to close safely the mission before the spacecraft re-entry. On Jan 18, 2024 the AGILE satellite stopped the scientific activities after almost 17 years of operations and, on Feb 14, 2024 re-entered uncontrolled into the atmosphere.

During the entire mission the ASI Broglio Space Center (BSC) in Malindi supported the AGILE mission through its Ground Stations for: data downlink and uplink communication channels; technical support for the development of AGILE-related activity that requires the use of the ground station facility at the BSC for all TT&C activity; engineering support in the development of SW for station upgrades; station readiness activities between the GS BSC and the AGILE Mission Operations Centre-launch in addition to the provision of support personnel.

## 4. The industrial contribution to the AGILE mission

Prominent companies of the Italian space industry participated in the development and mission management in orbit, including OHB Italia (former Carlo Gavazzi Space), Thales Alenia Space - Italia (former LABEN) and Leonardo (former Galileo

Avionica), Rheinmetall Italia (former Oerlikon Contraves), Telespazio, and Mipot.

The AGILE spacecraft has been developed by OHB Italia as prime contractor and by Rheinmetall Italia. The spacecraft has been designed to obtain a very good compromise between the mission technical requirements and the stringent volume and weight constraints.

The overall configuration has a hexagonal shape (for volume optimization and heat dissipation requirements) and is divided in two modules:
• the "Payload module", in the upper part of the spacecraft, hosting the instrument and its electronics, including the star sensors and GPS;
• the "service module" (i.e. the satellite platform) hosting the vital control units and in charge to provide the AGILE Payload with the required services for its operations.

In particular, the platform provides to the payload, for all the mission lifetime, power supply, communication with ground, attitude control and thermal control capability. All these functionalities are managed by the On Board Data Handling (OBDH) Subsystem that takes also care of the on board monitoring and control activities.

The electrical power for the Satellite is produced by a fixed solar panel of about 2 $m^2$ dimensions, equipped with triple junction GaAs cells. To supply power during eclipse periods and during the attitude acquisition phase, the power S/S includes a re-chargeable Lithium-Ion battery with a capacity of 33 Ah.

The battery charging activity and the power conversions and distribution to the satellite users is provided by a dedicated Power box controlled by the OBDH through dedicated software.

The communications with ground are ensured by an S band transceiver that contacts the ASI ground station located in Malindi (Kenia).

During the visibility periods the S band transmitter downloads the payload data, previously stored in the OBDH mass memory, with a net data rate of 500 Kbps.

The S band transceiver is also used to upload from ground the configuration telecommands necessary to the satellite operations.

The Attitude Control Subsystem (ACS) uses a set of sensors and actuators controlled by a dedicated software running on the on OBDH computer to guarantee the required attitude pointing in all the mission phases.

After the launcher separation, the ACS was able to acquire, within few orbits, a Sun pointing attitude to guarantee the necessary power generation from the solar array. Once acquired, the Sun pointing attitude is maintained for the whole mission using two different control algorithms: the Sun pointing spinning attitude, that make use of a reduced set of sensors and actuators, allowing to obtain a continuous sky survey and the

"fine" Sun pointing attitude (used the first part of the mission), that ensure a pointing accuracy better of 1 degree and an attitude stability better of 0.1 degrees/s providing the possibility to manoeuvre the payload instrument, while keeping the Sun pointing direction, in order to select the section of sky to be observed with the possibility to perform fast re-pointing towards active sources.

The Satellite thermal control is of passive type. It is realised mainly by placing the most dissipating devices on radiating surfaces. A proper material selection allows obtaining the desired temperature variations where all devices temperature limits fall in.

A set of heaters, controlled by thermostats or by SW, are used on critical equipment (in particular, the Li-Ion battery and the Payload) to guarantee the respect of the allowed temperature ranges.

To guarantee the required reliability for the whole mission lifetime, the main platform equipment has a full redundancy. The redundancy management is mainly performed autonomously by the OBDH SW in order to provide the required autonomy to recover possible failure that can be experienced during the mission.

## 5. The AGILE Ground Segment architecture and the data center at SSDC

The multi-mission ASI Space Science Data Center (SSDC) was established in late 2016 resting on the legacy of the ASI Science Data Center (ASDC), which in turn was built in 2000, on the experience of the BeppoSAX Science Data Center from the early 1990s. The SSDC is staffed by personnel from ASI, INAF and INFN, with an Information and Communication Technology (ICT) support provided by industrial partners (currently Telespazio and Serco).

The AGILE Data Center (ADC) is part of SSDC, and it acts both as a Science Operation Center (SOC) and as a Science Data Center (SDC) of the mission [10]. The AGILE ground segment architecture is schematically represented in Fig. 8.

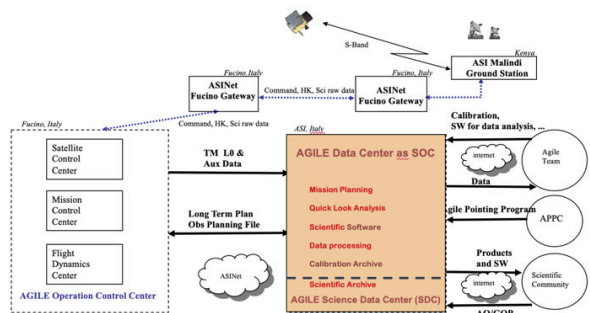

Fig. 8: A schematic view of the AGILE ground segment architecture.

The main ADC activities and responsibilities include:

- development and implementation of data processing systems (pipelines);
- development and maintenance of relational databases and data archives;
- real time data acquisition and monitoring;
- alert systems;
- processing (+reprocessing), from raw binary telemetry data (LV0) to scientific data products (LV3);
- data archiving and standardization;
- software tools development to grant optimal diffusion, accessibility and usability of scientific data according to FAIR principles.

The AGILE Science Program has been open for guest investigations on a competitive basis during the first four mission observation Cycles (2007-2011).

The ADC managed the Guest Observer (GO) Program of the AGILE mission with four Announcements of Opportunity (AO) issued by ASI.

Starting from October 2015, the standard one-year proprietary period requirement was eliminated, and all AGILE-GRID data have been published as soon as they were processed and validated. The ADC is responsible for the standardization, inclusion, distribution and maintenance of public scientific data in the large SSDC interactive multi-mission archive[1].

The AGILE scientific alert system was a distributed system, combining the ADC quick-look analysis with the AGILE Science Alert System developed by the AGILE team [11], generating alerts both on on a daily timescale and on a contact-by-contact basis. These alerts were sent via SMS, e-mail, and through the notification system of the dedicated App "AGILEScience" for smartphones and tablets [12].

The system allowed the AGILE team to perform a full data reduction and the preliminary quick-look scientific analysis as quickly as 25–30 minutes after the telemetry download from the spacecraft, a record for a gamma-ray mission.

## 6. AGILE main scientific results and the mission legacy

We briefly present here a selection of the main AGILE science highlights after more than sixteen years of operations, including some recent updates.

### 6.1 Discovery of Crab Nebula fast variability in gamma-rays

The Crab Nebula is the remnant of a supernova explosion witnessed by Chinese astronomers in 1054, energized by a powerful and fast rotating pulsar present in its inner part. The electomagnetic emission originating from the nebula has been considered to be essentially stable, and it is often used as a calibration source in X-ray astronomy. The surprising discovery by AGILE of variable gamma-ray emission above 100 MeV over a daily timescale from the Crab Nebula announced in Sept. 2010 started a new era of investigation of the Crab system [13]. The 2012 Bruno Rossi International Prize has been awarded to the PI, Marco Tavani, and the AGILE team for this important and unexpected discovery, which was then confirmed one day later by the NASA Fermi satellite [14].

Gamma-ray data provide evidence for particle acceleration mechanisms in nebular shock regions more efficient than previously expected from theoretical models.

### 6.2 First evidence of cosmic-ray acceleration from supernova remnants

Understanding the origin of galactic cosmic rays is one of the key topics in high-energy astrophysics and Supernova Remnants (SNR) are considered an ideal system to study cosmic-ray (hadronic) acceleration.

The breakthrough in this context was obtained by AGILE in the case of the intermediate-aged SNR W44 [15,16]. The AGILE-GRID data, which reach optimal sensitivity just at energies between 50 MeV and a few GeV, allowed to disentangle for the first time the signature of hadronic gamma-ray emission from pion decay in proton-proton collisions from that of electron Bremsstrahlung and inverse Compton processes. The observed AGILE gamma-ray spectrum is consistent with pion decay processes, showing a clear indication of the so-called "pion bump" in the 50–100MeV energy range. The modelling required a detailed analysis of the electron leptonic processes made possible by the co-spatial radio shell detection.

### 6.3 Detection of transient gamma-ray emission from microquasars

The study of microquasars was one of the key goals of the AGILE scientific program since the beginning of the mission. Microquasars are Galactic binary systems consisting of a neutron star or stellar-mass black hole accreting gas from a companion star, producing relativistic radio jets. Before AGILE, there was no systematic evidence for emission at gamma-ray energies from microquasar.

Three microquasars were detected by AGILE at energues above 100 MeV in the Cygnus region with variable gamma-ray emission: Cyg X-1 [17], Cyg X-3 [18], and V404 Cygni [19]. In the case of the Cyg X-3 source, AGILE detected for the first time several gamma-ray flares above 100 MeV following a clear repetitive pattern, either in coincidence with low hard X-ray fluxes or during transitions from low to high hard

---

[1] https://www.ssdc.asi.it/mmia/index.php?mission=agilemmia

X-ray fluxes, and usually appearing before major radio flares [18].

The important AGILE discovery shows that gamma-ray flares happen at prominent minima of the hard X-ray flux, indicating a very efficient particle acceleration mechanism leading to gamma-ray emission which occurs in coincidence with the launch of relativistic jets.

### 6.4 AGILE GRB and other transients: GW, neutrinos, FRBs, TGFs and Solar flares

Since the beginning of its mission AGILE detected Gamma-ray Bursts (GRB), extragalactic high-energy impulsive events. While Super-AGILE was capable of localizing about 1 GRB/month to within a few arcminutes, a larger number of GRBs have been detected by non-imaging 4π detector MCAL. An updated catalog of more than 500 GRBs detected by MCAL over a period of 13 years, from November 2007 to November 2020 has been recently published in [20]. A few % of the GRB samples may show a delayed high-energy component detectable by the AGILE-GRID above 100 MeV. Among the exceptional events observed by AGILE, the long-duration GRB 221009A needs to be mentioned, as it was the brightest and most energetic GRB ever recorded (the brightest of all time or the "BOAT") [21].

The AGILE space mission, with its fast ground segment alert system and its unique observing capability to cover about 80% of the sky in ~7 min in the so-called "spinning observing mode", also provided crucial contribution in follow-up observations of multiwavelength and multimessenger transients, such as gravitational wave events (GW), cosmic neutrinos, and fast radio bursts (FRB). AGILE also produced important results on terrestrial gamma-ray flashes (TGF) and solar flares, with the publication of dedicated catalogs.

For more information on the AGILE scientific results, we refer to two recent reviews [22,23].

## 7. Conclusions

During almost 17 years, AGILE contributed in fundamental ways to high-energy astrophysics, cosmic-ray physics, solar physics and to the study of terrestrial gamma-ray flashes.

With AGILE's re-entry, the in-orbit operational phase ended, but a new phase of scientific work on the satellite legacy data archive opens. AGILE archives and catalogs are available to the community through the ASI SSDC. AGILE-GRID data can be freely downloaded and analysed with the open-source Python software package Agilepy [2] [24], and/or SSDC AGILE-LV3 online data analysis tool [3] [25].

---

[2] https://agilepy.readthedocs.io/en/1.6.4/
[3] https://www.ssdc.asi.it/mmia/index.php?mission=agilelv3mmia

The AGILE successful experience also constitutes a valuable heritage for new and future high-energy missions such as, among others, COSI, Gamma-FLASH and CTAO.

### Acknowledgements

This paper in written on behalf of all the AGILE Collaboration members and of the team of the AGILE Data Center (ADC) at SSDC. We acknowledge the effort of ASI and industry personnel in operating the ASI ground station in Malindi (Kenya), the Telespazio Mission Control Center at Fucino, and the data processing done at the ADC in Rome.

The ADC operated in close relationship with the Telespazio (TPZ) Mission Operation Center at Fucino (L'Aquila), Italy, in particular with the spacecraft operations manager (SOM) P. Tempesta.

We also acknowledge the AGILE Mission Board, composed of: the PI of the AGILE Mission M. Tavani, the Co-PI G. Barbiellini, the ASI Project Scientist P. Giommi, the current ASI Mission Director G. Valentini, as well as the former Mission Directors: L. Salotti (Apr 2007 - Sep 2010), G. Valentini (Sep 2010 - Jan 2015), F. D'Amico (Jan 2015 - Jun 2023), and the ASI representative E. Tommasi di Vignano.

We would like to acknowledge the financial support of ASI under contract to INAF: ASI 2014-049-R.0 dedicated to SSDC.

The scientific research carried out for the project has been partially supported under the grants ASI-I/R/045/04, ASI-I/089/06/0, and ASI-I/028/12/0 and subsequent addenda. ASI-INAF Program Manager: E. Cavazzuti.

### References

[1] M. Tavani. M. Tavani, G. Barbiellini, A. Argan et al., The AGILE mission. Astron. Astrophys. 502, 995, doi:10.1051/0004-6361/200810527

[2] Argan, A., Tavani, M., and Trois, A., The on-board data processing of the AGILE satellite, Rendiconti Lincei. Scienze Fisiche e Naturali, vol. 30, pp. 199–205, 2019. doi:10.1007/s12210-019-00846-0.

[3] G. Barbiellini, M. Boezio, M. Candusso et al., A wide aperture telescope for high energy gamma-rays detection. Nucl. Phys. B Proc. Suppl. 43, 253 (1995). doi:10.1016/0920-5632(95)00484-Q

[4] M. Prest, G. Barbiellini, G. Bordignon, et al., The AGILE silicon tracker: an innovative gamma-ray instrument for space. Nucl. Instrum. Methods Phys. Res. A 501, 280 (2003). doi:10.1016/S0168-9002(02)02047-8

[5] G. Barbiellini, G. Fedel, F. Liello et al., The AGILE silicon tracker: testbeam results of the prototype silicon detector. Nucl. Instrum. Methods Phys. Res.

A 490, 146 (2002). doi: 10.1016/S0168-9002(02)01062-8

[6] C. Labanti, A. Argan, A. Bulgarelli et al., The Mini-Calorimeter detector for the AGILE mission. Nucl. Phys. B Proc. Suppl. 150, 34 (2006). doi:10.1016/j.nuclphysbps.2004.06.002

[7] C. Labanti, M. Marisaldi, F. Fuschino et al., Design and construction of the Mini-Calorimeter of the AGILE satellite. Nucl. Instrum. Methods Phys. Res. A 598, 470 (2009). doi:10.1016/j.nima.2008.09.021

[8] M. Feroci, E. Costa, P. Soffitta et al., SuperAGILE: the hard X-ray imager for the AGILE space mission. Nucl. Instrum. Methods Phys. Res. A 581, 728 (2007). doi:10.1016/j.nima.2007.07.147

[9] F. Perotti, M. Fiorini, S. Incorvaia et al., The AGILE anticoincidence detector. Nucl. Instrum. Methods Phys. Res. A 556, 228 (2006). doi:10.1016/j.nima.2005.10.016

[10] C. Pittori and the AGILE-SSDC Team, The AGILE data center and its legacy, Rendiconti Lincei. Scienze Fisiche e Naturali, vol. 30, pp. 217–223, 2019. doi:10.1007/s12210-019-00857-x

[11] A. Bulgarelli et al., The AGILE Alert System for Gamma-Ray Transients, ApJ, 781, 19 (2014)

[12] N. Parmiggiani, A. Bulgarelli, M. Tavani et al., The AGILEScience mobile application for the AGILE space mission, Astron. and Comp., vol. 48, Art. no. 100849, 2024. doi:10.1016/j.ascom.2024.100849

[13] M. Tavani, A. Bulgarelli, V. Vittorini et al., Discovery of powerful gamma-ray flares from the crab nebula. Science 331, 736 (2011a). doi:10.1126/science.1200083

[14] A.A. Abdo, M. Ackermann, M. Ajello et al., Gamma-ray flares from the crab nebula. Science 331, 739 (2011). doi:10.1126/science.1199705

[15] A. Giuliani, M. Cardillo, M. Tavani et al., Neutral pion emission from accelerated protons in the supernova remnant W44. Astrophys. J. Lett. 742, L30 (2011). doi:10.1088/2041-8205/742/2/L30

[16] M. Cardillo, M. Tavani, A. Giuliani et al., The supernova remnant W44: confirmations and challenges for cosmic-ray acceleration. Astron. Astrophys. 565, A74 (2014). doi: 10.1051/0004-6361/201322685

[17] S. Sabatini, M. Tavani, E. Striani, et al., Episodic Transient Gamma-ray Emission from the Microquasar Cygnus X-1, Astrophys. J. Lett. 2010, 712, L10–L15. doi:10.1088/2041-8205/712/1/L10

[18] M. Tavani, A. Bulgarelli, G. Piano et al., Extreme particle acceleration in the microquasar CygnusX-3. Nature 462, 620 (2009b). doi:10.1038/nature08578

[19] G. Piano, P. Munar-Adrover, F. Verrecchia, et al., High-energy Gamma-ray Activity from V404 Cygni Detected by AGILE during the 2015 June Outburst. Astrophys. J. 2017, 839, 84. doi:10.3847/1538-4357/aa6796

[20] A. Ursi et al., The second AGILE-MCAL GRB catalog, Astrophys. J. 925, 152 (2022). doi:/10.3847/1538-4357/ac3df7

[21] M. Tavani, G. Piano, A. Bulgarelli, et al., AGILE Gamma-ray Detection of the Exceptional GRB 221009A, Astrophys. J. Lett.2023, 956, L23. doi:10.3847/2041-8213/acfaff

[22] M. Tavani, C. Pittori and F. Longo, The AGILE Mission and Its Scientific Results, (2023), Handbook of X-ray and Gamma-ray Astrophysics https://link.springer.com/referenceworkentry/10.1007/978-981-16-4544-0_57-1

[23] S. Vercellone, C. Pittori and M. Tavani, Scientific Highlights of the AGILE Gamma-ray Mission, (2024), Universe doi:10.3390/universe10040153

[24] A. Bulgarelli, L. Baroncelli, A. Addis et al., Agilepy: A Python Framework for AGILE Data Analysis, Astronomical Data Analysis Software and Systems XXX, 2022, vol. 532, p. 509. doi:10.48550/arXiv.2105.08474.

[25] F. Lucarelli and C. Pittori, The AGILE Gamma-Ray Legacy Archive and the User-Friendly AGILE-LV3 Web Tool Integrated in the ASI-SSDC MWL Environment, Astronomical Data Analysis Software and Systems XXIX, 2020, vol. 527, p. 33.